\documentclass[preprint,12pt]{elsarticle}

\usepackage{amsmath, amssymb, amsthm}
\usepackage{color}
\usepackage{latexsym}
\usepackage{indentfirst}
\usepackage{graphicx}
\usepackage{placeins}
\usepackage{booktabs}
\usepackage{algorithm}
\usepackage{algorithmic}
\usepackage{multirow}
\usepackage{subfig}
\usepackage{bbm}

\theoremstyle{plain}

\theoremstyle{definition}

\theoremstyle{remark}

\newcommand{\etc}{\textit{etc}{}}
\newcommand{\etal}{\textit{et al}{}}
\newcommand{\ie}{\textit{i.e.}{~}}
\newcommand{\eg}{\textit{e.g.}{~}}

\newcommand{\abs}[1]{\left\lvert#1\right\rvert}
\newcommand{\norm}[1]{\left\lVert#1\right\rVert}

\newcommand{\barint}{\kern4pt \raise3.4pt\hbox{\vrule height.6pt
    width7pt} \kern-11pt \int}

\newcommand{\bvec}[1]{\mathbf{#1}}

\newcommand{\vp}{\bvec{p}}
\newcommand{\vr}{\bvec{r}}
\newcommand{\vL}{\bvec{L}}
\newcommand{\vS}{\bvec{S}}

\newcommand{\vsigma}{\mbox{\boldmath$\sigma$}}
\newcommand{\vm}{\bvec{m}}
\newcommand{\vB}{\bvec{B}}
\newcommand{\vI}{\bvec{I}}
\newcommand{\vH}{\bvec{H}}
\newcommand{\vV}{\bvec{V}}
\newcommand{\vzero}{\bvec{0}}

\def\dif{\mathrm{d}}
\def\mi{\mathbbm{i}}

\journal{Journal of Computational Physics}

\begin{document}

\begin{frontmatter}

\title{Efficient iterative method for solving the Dirac-Kohn-Sham density
functional theory}
\author[ll]{Lin Lin}
\ead{linlin@lbl.gov}

\author[ss]{Sihong Shao\corref{cor}}
\ead{sihong@math.pku.edu.cn}

\author[we]{Weinan E}
\ead{weinan@math.princeton.edu}

\cortext[cor]{Corresponding author.}

\address[ll]{Computational Research Division, Lawrence
Berkeley National Laboratory, Berkeley, CA 94720, USA}

\address[ss]{LMAM and School of Mathematical Sciences, Peking University,
Beijing 100871, China}

\address[we]{Department of Mathematics and PACM, Princeton University,
Princeton, NJ 08544, USA; Beijing International Center for Mathematical
Research, Peking University, Beijing 100871, China}

\begin{abstract}
We present for the first time an efficient iterative method to directly
solve the four-component Dirac-Kohn-Sham (DKS) density functional theory.
Due to the existence of the negative energy continuum in the DKS
operator, the existing iterative techniques for solving the Kohn-Sham
systems cannot be efficiently applied to solve the DKS
systems.  The key component of our method is a novel filtering step (F)
which acts as a preconditioner in the framework of the locally optimal
block preconditioned conjugate gradient (LOBPCG) method.  The resulting
method, dubbed the LOBPCG-F method, is able to compute the desired
eigenvalues and eigenvectors in the positive energy band without
computing any state in the negative energy band.  The LOBPCG-F method
introduces mild extra cost compared to the standard LOBPCG method and
can be easily implemented.  We demonstrate our method in the
pseudopotential framework with a planewave basis set which naturally
satisfies the kinetic balance prescription.  Numerical results for
Pt$_{2}$, Au$_{2}$, TlF, and Bi$_{2}$Se$_{3}$ indicate that the LOBPCG-F
method is a robust and efficient method for investigating the
relativistic effect in systems containing heavy elements.
\end{abstract}


\end{frontmatter}

\section{Introduction}
\label{sec:intro}

The electron, as an elementary particle, has spin and charge, and
further acquires an angular momentum quantum number corresponding to a
quantized atomic orbital when binded to the atomic nucleus.  The
importance of the spin-orbit coupling (SOC) effect in semiconductors and
other materials have been extensively explored in quantum physics and
quantum chemistry.  For example, SOC causes shifts in the atomic energy
level of an electron~\cite{bk:Griffiths1995}, and leads to protected metallic
surface or edge states as a consequence of the topology of the bulk
electronic wave functions~\cite{KaneMele2005}.
As a typical relativistic effect,
SOC has magnitude of the order $Z^4\alpha^{-2}$ for a hydrogen
like atom~\cite{bk:AtkinsFriedman2005},
where $Z$ is the nuclear charge,
$\alpha=c^{-1}$ in the atomic unit is the fine structure constant and
$c$ is the speed of light.  For systems containing heavy elements with
large $Z$, the nonrelativistic Schr\"{o}dinger type equations such as
the Kohn-Sham (KS) density functional
theory~\cite{HohenbergKohn1964,KohnSham1965} leads to large error \cite{Pyykko1988,Pyykko2012}.
The extension of the density functional theory is not straightforward as
quantum electrodynamics has to been used for charged particles in which
complicated renormalization is necessary to get finite expressions
for charge, energy, \etc~\cite{Engel2002}.
The relativistic density
functional theory, first laid out by Rajagopal and Callaway~\cite{RajagopalCallaway1973},
can be rigorously derived from quantum
electrodynamics.  However, the resulting equations are too complicated
to solve in practice and proper renormalization has to be performed to
eliminate the divergent terms.  The frequently-used working equations
are the four-component Dirac-Kohn-Sham (DKS) equations derived by
Rajagopal~\cite{Rajagopal1978} and independently by MacDonald and Vosko~\cite{MacDonaldVosko1979}
after making several physically reasonable approximations.
The extension to the time-dependent scenario can be found in \eg \cite{Rajagopal1994,Rajagopal1996}.

\begin{figure}[htpb]
  \begin{center}
    \includegraphics[width=0.6\textwidth]{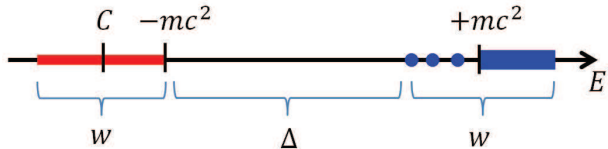}
  \end{center}
	\caption{A sketch of the spectrum of the DKS operator. For explanations see the text.}
  \label{fig:schemeLOBPCG}
\end{figure}

Mathematically, the DKS operator is fundamentally different from the KS operator.
The spectrum of the DKS operator,
sketched in Fig.~\ref{fig:schemeLOBPCG}, consists of the point
spectrum (blue dots), the positive energy continuum (thick blue line
from $+mc^2$ to $+\infty$) and the negative energy continuum (thin red
line from $-\infty$ to $-mc^2$), and is therefore unbounded from blow~\cite{bk:Thaller1992}.
Meanwhile the spectrum of the KS operator does not contain any negative
energy and is bounded from below.
Here $m$ is the electron mass and we have $m=1$ and $c\approx137$ in atomic unit.
When solving the DKS equations
numerically, the DKS operator will be discretized by a basis set of
finite size, and both the positive energy continuum and the negative
energy continuum will be truncated before infinity is reached.
Hereafter we refer to the set of eigenvalues of the
discretized DKS operator in the point spectrum and the positive energy
continuum as the \textit{positive energy band}, and the set of
eigenvalues of the discretized DKS operator in the negative energy
continuum as the \textit{negative energy band}.
There are approximately
equal number of eigenvalues in the positive and the negative energy
bands separated by a large gap, which is also called the forbidden
region with its width denoted by $\Delta$. We denote by $w$ the maximum of the widths of the
positive and the negative energy band. The desired eigenvalues and
eigenvectors in the positive energy band are usually contained in the
point spectrum or in very rare cases cover the entire point spectrum and
a very few low lying states in the positive energy continuum.

Attempts to solve the DKS equations are plagued by the
so-called ``variational collapse''~\cite{SchwarzWallmeier1982},
which specifically means two difficulties.
The first difficulty is the so-called spectral pollution -- the appearance of spurious
eigenvalues which are the limiting points of the eigenvalues of the
discretized DKS operator as the basis set becomes complete, but are not
the eigenvalues of the true DKS operator.  They often lie deeper than
the desired solutions or may even be degenerate with them.  The second
difficulty is
the occurrence of the spurious eigenvalues in the forbidden region when
solving the discretized DKS operator using inappropriate numerical
schemes.
The reason for such collapse is that the spectrum of the DKS operator
cannot be defined variationally~\cite{bk:Thaller1992}.  Several
prescriptions to avoid variational collapse were summarized and analyzed
by Kutzelnigg~\cite{Kutzelnigg1984} as well as Lewin and S{\'e}r{\'e}~\cite{LewinSere2010}.
Among all the prescriptions, the kinetic balance
prescription~\cite{IshikawaBinningSando1983,StantonHavriliak1984,DyallFagri1990}
addresses the first difficulty most successfully from practical point of
view, and serves as the foundation for most successful attempts to solve
the DKS equations via finite dimensional matrix eigenvalue problems.
Many effective numerical implementations
for performing four-component DKS calculations with the
localized basis sets (\eg~Gaussian type orbitals and atomic orbitals)
have been presented by quantum chemists in the last four decades,
including the DVM scheme~\cite{Rosen1975,Ellis1984}, the BDF package~\cite{Liu1997},
the program of Fricke and coworkers~\cite{Varga2000},
the REL4D module in Utchem~\cite{Yanai2001},
the DKS module in DIRAC~\cite{Saue2002} and the BERTHA code~\cite{Quiney2002}.
In all these methods, the finite dimensional
DKS matrix is diagonalized directly as a dense matrix, which
computes all the eigenvalues and eigenvectors in both positive and
the negative energy bands.  The direct diagonalization methods avoid the
second difficulty in the variational collapse originating from
inappropriate numerical schemes. However, these methods are
prohibitively expensive when the dimension of the discretized DKS
operator becomes large.

While the localized basis sets are widely used in the community of
quantum chemistry, the planewave basis set is more popular in the
community of condensed matter physics for the reason that systematic
convergence can be achieved by only increasing the kinetic energy
cutoff. The planewave basis set can be used as the only basis set in the
pseudopotential framework~\cite{TroullierMartins1991}, and
as an augmented basis set such as in the linearized augmented planewave
method (LAPW)~\cite{Andersen1975}.
In terms of the DKS density functional theory,
the planewave basis set automatically
satisfies the kinetic balance prescription and is free from the spectral
pollution \cite{LewinSere2010}.
However, compared to the localized basis set, the planewave
basis set leads to matrix eigenvalue problems of much larger size which
is impractical to be diagonalized directly as dense matrices.
Iterative methods have to be designed to solve the matrix eigenvalue
problems for the desired eigenvalues and eigenvectors contained in the positive
energy band.
There are
approximately the same number of eigenvalues in the positive and the
negative energy bands, and standard iterative techniques for solving the
KS systems cannot be efficiently applied without necessary
modification.  For example, the conjugate gradient
method~\cite{StichCarParrinelloEtAl1989,PayneTeterAllenEtAl1992},
the locally optimal block preconditioned conjugate gradient (LOBPCG)
method~\cite{Knyazev2001,KnyazevArgentatiLashukEtAl2007}, the RMM-DIIS
method~\cite{KresseFurthmuller1996} and the Lanczos
method~\cite{WuCanningSimonEtAl1999} need to evaluate all the states in
the negative energy band and therefore is prohibitively expensive; It is
difficult to design a set of efficient filtering polynomials as in the
Chebyshev filtering method~\cite{ZhouSaadTiagoEtAl2006},
since the separation between the occupied and unoccupied eigenvalues in the
positive energy band (usually in the order of $0.1$ au or smaller) is
much smaller compared to the width of the forbidden region
($\Delta\approx 37000$ au).
The spectral radius of the matrix resulting from the spectrum folding
technique~\cite{WangZunger1994} will
be in the order of $\Delta^2\approx 10^{9}$ for the DKS
equations, and the resulting positive definite eigenvalue problem cannot
be solved efficiently with iterative methods.

In this work, an efficient iterative method to directly solve the
four-component DKS density functional theory will be
presented for the first time.
The key component of our method is a filtering step (F) which can act as
a preconditioner in preconditioned iterative diagonalization methods.
In particular, our method is demonstrated based on the LOBPCG
method, and is therefore dubbed the LOBPCG-F method.  Compared to other
types of preconditioned conjugate gradient methods, the LOBPCG method
has been shown to be effective for evaluating a relatively large number
of eigenvalues and eigenvectors, and its efficiency has been illustrated
in large scale electronic structure calculation such as in
ABINIT~\cite{BottinLerouxKnyazevEtAl2008,Abinit1}.  However, for the DKS
systems, the standard LOBPCG method amplifies the error of the
eigenfunctions projected to the negative energy band in each iteration,
and therefore needs to evaluate all the eigenfunctions in the negative energy
band together with the desired eigenfunctions in the positive energy
band.  We illustrate that such error can be efficiently controlled by
the filtering step, and the desired eigenvalues and eigenvectors in the
positive energy band can be evaluated without computing any negative
energy state.  The filtering step only introduces $2$ extra
matrix-vector multiplication per iteration, and can be implemented
simply as a preconditioner with little coding effort.

We implement the LOBPCG-F method to solve the DKS
equations in the pseudopotential framework with a planewave basis set
(module DKS). To benchmark our numerical results,
we also implement the standard LOBPCG method for solving the KS density
functional theory (module KS), and for solving the two-component KS
density functional theory with SOC in the pseudopotential framework
(module KSSO).  We directly compare the total energies obtained from
the modules KS and KSSO with those obtained from the
corresponding modules in a popular electronic structure software
ABINIT~\cite{Abinit1}.  We apply the modules KS, KSSO and DKS to solve
systems including Pt$_{2}$, Au$_{2}$, TlF, and Bi$_{2}$Se$_{3}$.
The differences in the total free energy are less than $1$ meV
per atom compared to ABINIT for all systems under study.
Despite that the relativistic correction is relatively small for
valence-valence interaction, our numerical results for solving the DKS equations
indicate that the LOBPCG-F method is robust and efficient in studying
the relativistic effect in systems containing heavy elements.

All the discussion in the rest of the manuscript will be presented in
the pseudopotential framework in the context of orthogonal basis set
(the planewave basis set is orthogonal).  We remark that the LOBPCG-F
method can be used for all-electron calculation and for
non-orthogonal basis functions as well.  This is the case for the LAPW
method such as implemented in the WIEN2k
software~\cite{SchwarzBlahaMadsen2002}, and also for the localized basis
set given that the overlap matrix is not very ill-conditioned.

The paper is organized as follows. We briefly review the four-component
DKS density functional theory in
Section~\ref{sec:reldft-dks} and discuss in detail the variational
collapse and the choice of basis set.  We develop the LOBPCG-F algorithm
in Section~\ref{sec:lobpcgf}.  The numerical results are
presented with discussion in Section~\ref{sec:numericalresults}. The
paper is concluded in Section~\ref{sec:conclusion}.
Throughout the paper, the atomic units ($e=\hbar=1$) will be used unless
otherwise noted.

\section{DKS density functional theory}
\label{sec:reldft-dks}

We will now describe the main working equations -- the DKS equations
derived by Rajagopal \cite{Rajagopal1978} and independently
by MacDonald and Vosko \cite{MacDonaldVosko1979},
and refer to Engel~\cite{Engel2002} and van W\"{u}llen~\cite{Wullen2010}
for a elaborate and general discussion on
relativistic density functional theory which can be rigorously derived from
quantum electrodynamics.
Several related numerical issues are then discussed,
such as the variational collapse
and the choice of basis in solving the DKS matrix eigenvalue problems.

\subsection{DKS equations}

Under the no-pair and electrostatic approximations~\cite{Engel2002,Wullen2010},
the total energy of
an interacting $n$-electron system corresponding to the Dirac-Coulomb
Hamiltonian takes the form
\begin{equation}\label{totalenergy}
E[\rho,\vm] =
T[\rho,\vm]
+\int V_{\text{ext}}(\vr) \rho(\vr)\dif \vr
+\frac12\int\frac{\rho(\vr_1)\rho(\vr_2)}{|\vr_1-\vr_2|}\dif\vr_1\dif\vr_2
+E_{\text{xc}}[\rho,\vm],
\end{equation}
where $\rho(\vr)$ is the electron density, $\vm(\vr)$ is the spin
density,
and $V_{\text{ext}}(\vr)$ is the nuclear attractive potential.
The first term $T$ is the kinetic energy of
the noninteracting reference system suggested by Kohn and Sham
\cite{KohnSham1965}, the third term is the electrostatic energy, and
$E_{\text{xc}}$ is the exchange-correlation functional containing both
the difference between the true many body electron-electron repulsive energy
and the electrostatic energy, and the difference between the true many
body kinetic energy and the kinetic energy of the noninteracting
reference system.
This factious system, which could be represented
by a single Slater determinant in terms of one-electron
spinors $\{\psi_i\}$ corresponding to the energy levels $\{\epsilon_i\}$,
has the same electron density as the interacting many body system. In consequence,
we could use $\{\psi_i\}$
to evaluate the kinetic energy and the densities
\begin{align}
T&= \sum_{i}^{occ} \psi_i(\vr)^\dag h_{\text{D}} \psi(\vr),\quad
h_{\text{D}} = \left(
\begin{matrix}
mc^2\vI_2 & c \vsigma \cdot \vp \\
c \vsigma \cdot \vp & -mc^2\vI_2
\end{matrix}
\right), \\
\rho(\vr) &= \sum_i^{occ} \psi_i(\vr)^\dag \psi_i(\vr),\quad
\vm(\vr) = \sum_i^{occ} \psi_i(\vr)^\dag
\left(
\begin{matrix}
\vsigma & \vzero_2\\
\vzero_2 & -\vsigma
\end{matrix}
\right)
\psi_i(\vr),\label{density}
\end{align}
where $h_{\text{D}}$ is the Dirac operator corresponding to free
particles,
$\vp = -\mi \nabla$ is the momentum operator,
$\vI_n$ and $\vzero_n$ are the $n\times n$ unit and null matrices,
$\vsigma=(\sigma_x,\sigma_y,\sigma_z)$ is the vector of
the Pauli spin matrices, and the each spinor
$\psi_i$ is a complex vector-valued function: $\mathbb{R}^3\rightarrow \mathbb{C}^4$, which are often rewritten as $\psi_i = (\phi_i,\chi_i)^T$
with $\phi_i,\chi_i$ being functions: $\mathbb{R}^3\rightarrow \mathbb{C}^2$. In what follows we will often refer to the two-spinor
$\phi_i$ (resp. $\chi_i$) as the large (resp. small) component of the
four-spinor $\psi_i$. Here the summations are only restricted to the
occupied states in the positive energy band.

The Euler-Lagrange equation with respect to $E[\rho,\vm]$ gives rise to
the DKS equation 
\begin{equation}\label{DKS}
\left[
\left(
\begin{matrix}
\vsigma \cdot \vB_{\text{xc}} + mc^2\vI_2 & c \vsigma \cdot \vp \\
c \vsigma \cdot \vp & -\vsigma \cdot \vB_{\text{xc}}-mc^2\vI_2
\end{matrix}
\right) + V_{\text{hxc}}\vI_4 + V_{\text{ext}}\vI_4\right]\left(
\begin{matrix}
\phi_i \\
\chi_i
\end{matrix}
\right) =
\left(
\begin{matrix}
\phi_i \\
\chi_i
\end{matrix}
\right) \epsilon_i,
\end{equation}
where
\begin{align}
V_{\text{hxc}}(\vr) &= \int\frac{\rho(\vr_1)}{|\vr-\vr_1|}\dif\vr_1
+ \frac{\delta E_{\text{xc}}}{\delta\rho}(\vr),\\
\vB_{\text{xc}}(\vr) &= \frac{\delta E_{\text{xc}}}{\delta\vm}(\vr).
\end{align}
Here for simplicity the exchange-correlation functional under local
density approximation (LDA)~\cite{CeperleyAlder1980,PerdewZunger1981} is
used. Our discussion below is not restricted to the LDA approximation.

Ideally all-electron calculations should be performed to obtain accurate
results, but with large computational cost in order to resolve the sharp
gradient of core orbitals and the oscillation of valance orbitals in the
neighborhood of nuclei.  A more practical way striking the balance
between accuracy and efficiency is to employ effective core
potentials~\cite{DolgCao2012}, in which the core electrons are frozen
and the valance-only problem is solved.  The pseudopotential technique
is one branch of effective core potential approaches.  One should be
careful in choosing pseudopotential for a relativistic Hamiltonian such
as DKS, since most of the existing pseudopotentials are generated for
the nonrelativistic Hamiltonian.  The mismatch between the
pseudopotential and the relativistic Hamiltonian introduce possibly
double counting of the relativistic effect \cite{IshikawaMalli1981}.  We
adopt here the widely known HGH
pseudopotential~\cite{HartwigsenGoedeckerHutter1998} which includes the
scalar relativistic effect and the spin-orbit coupling effect of the
core electrons by construction, and can be specified by a very small
number of parameters due to its dual-space Gaussian form.  Rigorously
speaking, the HGH pseudopotential may still suffer from double counting
of the relativistic effect, since the parameters are optimized using the
non-relativisitic equations. On the other hand, the numerical method
developed in this paper does not depend on the specific choice of
parameters of the HGH pseudopotential, and the numerical results can be
readily improved when the HGH pseudopotential is reparametrized for the
DKS calculation.
The HGH pseudopotential consists of three parts
\begin{equation}
\vV_{\text{HGH}}(\vr,\vr')=V_{\text{loc}}(\vr)\delta(\vr-\vr')\vI_4 +
  \sum_{l}\left(V_{l}(\vr,\vr') \vI_4 + \Delta V_{l}^{\text{SO}}(\vr,\vr')
	\vL'\cdot\vS\right),
	\label{eqn:HGH}
\end{equation}
where the subscript $l$ is the angular momentum number,
$V_{\text{loc}}$, $V_l$, and $\Delta V_{l}^{\text{SO}}$ represent in the
order the local contribution, the nonlocal contribution
and the SOC effect of the HGH pseudopotential, of which the formulas can
be found in \cite{HartwigsenGoedeckerHutter1998} and thus are skipped
here to save space. $\delta(\vr)$ is the Dirac delta function,
$\vL'$ is the angular momentum at position $\vr'$,
and $
\vS=\frac12\left(
\begin{matrix}
\vsigma & \vzero_2 \\
\vzero_2      & \vsigma
\end{matrix}\right)
$ is the spin operator.
The HGH pseudopotential $V_{\text{HGH}}$ is an integral operator on the spinors.
After denoting the first matrix operator of the LHS term of Eq.~\eqref{DKS} by $\vH_{0}$,
we finally arrive the DKS equation for valence electrons
\begin{equation}\label{vDKS}
\left[\vH_{0}  + V_{\text{hxc}}\vI_4\right]\psi_i(\vr) + \int \vV_{\text{HGH}}(\vr,\vr') \psi_i(\vr')\dif\vr' = \psi_i(\vr) \epsilon_i.
\end{equation}
The nuclear attraction term $V_{\text{ext}}$ in Eq.~\eqref{DKS} is replaced here
by the $\vV_{\text{HGH}}$ term that describes the electrostatic interaction
between the effective nuclei and valance electrons.
The DKS equations~\eqref{DKS} or~\eqref{vDKS} are usually solved with the
self-consistent field (SCF) iteration as in the case of the
nonrelativistic KS equations~\cite{Martin2004}.

\subsection{Variational collapse and kinetic balance}

As we have mentioned in Section~\ref{sec:intro}, the most salient
feature of the DKS operator is that its spectrum contains the negative
energy continuum and is therefore unbounded from below.  The desired
eigenvalues and eigenvectors corresponding to the occupied bound states
lie in the positive energy band (see Fig.~\ref{fig:schemeLOBPCG}),
\ie we are seeking for highly excited states above
all the negative energy states.
When solving the resulting matrix eigenvalue problems by expanding $\psi_i$ in Eq.~\eqref{vDKS}
with a finite basis set,
such feature of the DKS operator imposes a large obstacle,
\ie the variational collapse called by Schwarz and Wallmeier~\cite{SchwarzWallmeier1982},
in the sense that the eigenvalues of the discretized DKS operator may
not systematically converge to the eigenvalues of the true DKS operator
as the basis size increases.
This variational collapse, is imputed by Kutzelnigg~\cite{Kutzelnigg1984}
to the wrong nonrelativistic limit of the matrix representation of the Dirac operator
and several prescriptions have been proposed to avoid it.
The first strategy is to replace minimization procedures
by min-max
procedures~\cite{Talman1986,DattaDevaiah1988,DesclauxDolbeaultEstebanIndelicatoSere2003}.
The min-max methods find the minimum over the large component of the
spinors and the maximum over the small component of the spinors for the
DKS energy functional, and these methods are not used often practically in
electronic structure calculation.
The second strategy is the two-component relativistic theory which
seeks for operators which are bounded from below and agrees with the DKS operator
in the nonrelaticistic limit~\cite{LiShaoLiu2012}.  A very good discussion
of approaches made in this direction can be found in~\cite{Liu2010}.
The third strategy is to carefully choose the basis set
with which the discretized DKS operator is free of the spectral pollution.
Among all the strategies, the kinetic balance
method~\cite{IshikawaBinningSando1983,StantonHavriliak1984,DyallFagri1990},
falling into the third category,
is widely used by theoretical chemists~\cite{Liu2010}, and serves as
the basis of most successful attempts to solve matrix eigenvalue
equations based on finite-dimensional representations of the DKS
operator.
The feature of the kinetic balance is the one-to-one
correspondence of the basis sets that are used to expand the large and
small components, achieved via  a linear operator $\vsigma\cdot\vp$.
Specifically, the basis $\{f_{k},,k=1,\cdots,N\}$ for expanding the
small components are obtained directly from the basis
$\{g_k,k=1,\cdots,N\}$ for expanding the large components, according to
the so-called kinetic balance prescription
\begin{equation}\label{RKB}
f_k = \frac{1}{2mc} \vsigma\cdot\vp g_k, \quad k=1,\cdots,N.
\end{equation}
The use of the kinetic balance avoids the worst aspects of
variational collapse and could generate the desired eigenvalues and
eigenvectors in the positive energy band with good accuracy if a
sufficiently flexible basis set is employed
\cite{StantonHavriliak1984,DyallFagri1990,LewinSere2010,SunLiuKutzelnigg2011}.
The localized basis sets satisfying the kinetic balance prescription
\eqref{RKB} are often used in quantum chemistry and all the eigenvalues
and eigenvectors of the discretized DKS operator are obtained by the
direct diagonalization method including the negative energy band that
are not used in practice.

\subsection{Choice of basis set: Planewaves}

Compared to the standard localized basis sets used in the relativistic
quantum chemistry community, the planewave basis set has the advantage
that systematic convergence can be achieved by tuning one parameter --
the kinetic energy cutoff~\cite{Martin2004}.  Furthermore, the planewave
basis set automatically satisfies the kinetic balance
prescription~\eqref{RKB}, \ie  the space spanned by planewaves within a
certain cutoff is invariant when applied by the $\vsigma\cdot \vp$
operator, and therefore the planewave basis set is a good basis set for
both the large component and the small component. Namely, the planewave
basis set can avoid the spectral pollution in practice.  However, the
planewave basis set usually results in a much larger degrees of freedom
per atom (in the order of $500\sim 5000$ or more for standard norm
conserving pseudopotential~\cite{TroullierMartins1991} with only valence
electrons taken into consideration), which is much larger than the
degrees of freedom per atom used by localized basis set.  We remark that
the degrees of freedom used by the finite difference method and the finite
element method will also be much larger than the degrees of freedom used
by localized basis set, but it is less obvious how the kinetic balance
prescription will be enforced in these methods.

Due to the large matrix size after the planewave discretization, the
direct diagonalization method will not be applicable.  The discretized
Hamiltonian operator in \eqref{vDKS} generated by the planewave basis
set, hereafter also referred to as the Hamiltonian matrix, is dense in
both the Fourier space and the real space, which discourages the usage
of the shift-invert Lanczos method~\cite{ParlettSaad1987}, the contour
integral based spectrum slicing methods~\cite{Polizzi2009}, or the Fermi
operator expansion based low order scaling
algorithms~\cite{LinLuYingE2009,LinLuYingCarE:09,Ozaki2010}.  These
methods involve matrix inversion and become less competitive when the
matrix is dense.
Furthermore, techniques that directly eliminate the negative energy
band, such as exact two-component theory~\cite{Liu2010} involves direct
manipulation of the Hamiltonian matrix, which is again not possible to
handle in the case of the planewave basis set as a result of the large
matrix size.  Iterative methods have to be used to diagonalize the
discretized Hamiltonian operator, which will be discussed in detail in
Section~\ref{sec:lobpcgf}.

\section{An efficient iterative method for solving the DKS equations}
\label{sec:lobpcgf}

As mentioned in Section~\ref{sec:intro}, there are approximately the
same number of eigenvalues and eigenvectors in the positive and the negative energy
bands, separated by a large spectral gap with width $\Delta$ as shown in Fig~\ref{fig:schemeLOBPCG}.
Existing iterative techniques
for solving the KS equations cannot be efficiently applied to solve the
DKS equations without necessary modification.  In this section we
demonstrate  that the negative energy band of the DKS operator can be
efficiently eliminated using only $2$ additional matrix-vector
multiplications per iteration, based on the locally optimal block
preconditioned conjugate gradient (LOBPCG) method~\cite{Knyazev2001}.
The key component of
the proposed new method is an additional filtering (F) step, and we therefore
refer to our new method as the LOBPCG-F method.  We remark that a
similar filtering procedure can also be applied to other preconditioned
iterative solvers to eliminate the negative energy band, such as the
preconditioned Davidson type methods~\cite{SleijpenVanDerVorse2000}.

\subsection{LOBPCG-F algorithm}

\begin{algorithm}[htpb]
\begin{small}
	\begin{center}
		\begin{minipage}[t]{5in}
			\begin{tabular}{p{0.5in}p{4.5in}}
				{\bf Input}:  &  \begin{minipage}[t]{4.0in}
					Subroutine $Hx$ ($Tx$) to multiply the Hamiltonian
					(preconditioning) operator to a vector,
					the shift constant $C$, the normalization constant $\Omega$,
					the number of eigenvalues $n$, the initial guess of eigenvectors $X_0\in \mathbb{R}^{N\times
					n}$, the number of initial filtering steps $N_{\mathrm{init}}$, the
					maximum number of iterations $N_{\mathrm{iter}}$, the
					tolerance for convergence $\epsilon$.
				\end{minipage} \\
				{\bf Output}:  &  \begin{minipage}[t]{4.0in}
					$\Lambda=\mathrm{diag}(\lambda_{1},\ldots,\lambda_{n})$ where
					$\{\lambda_{i}\}_{i=1}^{n}$ are the lowest $n$ eigenvalues in the
					positive energy band,
					and $X\in \mathbb{R}^{N\times n}$ are the associated eigenvectors.
				\end{minipage}
			\end{tabular}
			\begin{algorithmic}[1]
				\FOR{$\underline{k=1,\ldots,N_{\mathrm{init}}}$}
				\STATE \underline{Set $X_{0}\gets f(H) X_{0}.$}
			\ENDFOR
			\STATE Solve $V$, $\Lambda_{1}\in \mathbb{R}^{n\times n}$ from the generalized eigenvalue
			problem $(X_{0}^{T}HX_{0}) V = (X_{0}^{T}X_{0}) V \Lambda_{1}$,
			with the diagonal elements of $\Lambda_{1}$ sorted in a
			non-decreasing order.
			\STATE Set $X_{1}\gets X_{0}V$.
			\STATE Compute the residual $R\gets HX_{1}-X_{1}\Lambda_{1}$.
			\STATE Set $P\gets []$ to be an empty matrix.
			\FOR{$k=1,\ldots,N_{\mathrm{iter}}$}
			\STATE Solve the preconditioned residual $Y$ from $TY=R$.
			\STATE \underline{Set $Y\gets f(H) Y$.}
			\STATE Construct a subspace $Z=[X_{k}, Y, P]$, $Z\in
			\mathbb{R}^{N\times n_{Z}}$ and $n_{Z}>n$.
			\STATE Solve $V, \Lambda_{k+1}\in \mathbb{R}^{n_Z\times n_Z}$ from the generalized
			eigenvalue problem
			$(Z^{T}HZ)V = (Z^{T}Z)V\Lambda_{k+1}$, with the diagonal elements
			of $\Lambda_{k+1}$ sorted in a non-decreasing order.
			\STATE Set $V\gets$ the first $n$ columns of $V$.
			\STATE Set $\Lambda_{k+1}\gets$ the upper-left $n\times n$ block of
			$\Lambda_{k+1}$.
			\STATE Set $X_{k+1}\gets Z V$.
			\STATE Compute the residual $R\gets HX_{k+1}-X_{k+1}\Lambda_{k+1}$.
			\IF{ the norm of each column of $R$ is less than $\epsilon$ }
			\STATE Exit the loop.
		\ENDIF
		\STATE Set $R\gets$ the columns of $R$ with norm larger than
		$\epsilon$.
		\STATE Set $V\gets$ the columns of $V$ corresponding to remaining
		columns of $R$.
		\STATE Set $P\gets [0,Y,P]V$.
	\ENDFOR
	\RETURN $X\gets X_{k+1}, \Lambda\gets \Lambda_{k+1}$.
\end{algorithmic}
 \end{minipage}
	\end{center}
\end{small}
  \caption{The LOBPCG-F algorithm for solving the DKS density
	functional theory. The underlined steps describe the new filtering
	step compared to the standard LOBPCG method. The filter is given in Eq.~\eqref{eqn:filter}.}
  \label{alg:LOBPCGF}
\end{algorithm}

The LOBPCG-F algorithm is described in Alg.~\ref{alg:LOBPCGF}.
It can be easily found there that
removing the new filtering steps underlined in Alg.~\ref{alg:LOBPCGF}
yields directly the standard LOBPCG algorithm as in~\cite{Knyazev2001}.
The matrices $H$, $T$ and $f(H)$ used in Alg.~\ref{alg:LOBPCGF} can be
defined in the ``matrix-free'' form, in the sense that they are defined
according to subroutines which only computes $Hx, Tx, f(H)x$ for any
vector $x$ without forming the matrix elements explicitly.
$C$ is a constant used as a shift in the filtering function,
which is chosen from the negative energy continuum as shown in Fig.~\ref{fig:schemeLOBPCG}.

We first demonstrate how the error of the eigenfunctions projected to
the negative energy band is propagated in the standard LOBPCG method.
To simplify the analysis,
we assume here that the preconditioner $T$ to be an identity
operator.  The qualitative result remains to be valid if standard
preconditioners in the electronic structure calculation are applied,
such as the preconditioner proposed by Teter \etal~\cite{PayneTeterAllenEtAl1992,TeterPayneAllan1989}.
Furthermore, we need a key assumption that
\begin{equation}
	\Delta \gg w,
	\label{eqn:assumptionDW}
\end{equation}
which is valid in the pseudopotential framework with the planewave basis set
for $\Delta\approx 37000$ and $w\approx 100$ hold there.
The foregoing assumption is also valid in the all-electron
calculation using the LAPW basis set, or using the localized basis set
given that the basis set remains well conditioned.
However, we remark that $w$ can be artificially large when the basis set is overcomplete
and then the assumption~\eqref{eqn:assumptionDW} will not be valid anymore.

The LOBPCG method computes the residual
$R=HX-X\Lambda$ once per
iteration (line $16$ in Alg.~\ref{alg:LOBPCGF}).
Denote by $x$ a given column of $X$,
$\lambda$ the corresponding eigenvalue in $\Lambda$,
and $r$ and $y$ the corresponding column in the
residual $R$ and the preconditioned residual $Y$, respectively.
We express $r$
and $x$ using the
eigen-decomposition as
\begin{equation}
	\begin{split}
	r = \sum_{i} \psi^{+}_{i}\tilde{r}^{+}_{i}
	   +\sum_{j} \psi^{-}_{j}\tilde{r}^{-}_{j},\\
	x = \sum_{i} \psi^{+}_{i}\tilde{x}^{+}_{i}
	   +\sum_{j} \psi^{-}_{j}\tilde{x}^{-}_{j},\\
	\end{split}
	\label{}
\end{equation}
Here $\psi^{+}_{i}$ is the spinor corresponding to $\epsilon^{+}_{i}$ in
the positive energy band, and $\psi^{-}_{j}$ the spinor corresponding
to $\epsilon^{-}_{j}$ in the negative energy band.   We further assume
that the error of the eigenfunctions projected to the negative energy
band, characerized by $\max_{j}|\tilde{x}^{-}_{j}|$,
is initially very small but
not yet vanishes.  Since $r=Hx-x\lambda$,  the coefficients
$\{\tilde{r}^{+}_{i}\}$, $\{\tilde{r}^{-}_{j}\}$
are related to the coefficients $\{\tilde{x}^{+}_{i}\}$,
$\{\tilde{x}^{-}_{j}\}$ according to
\begin{equation}
	\tilde{r}^{+}_{i}=(\epsilon^{+}_{i}-\lambda) \tilde{x}^{+}_{i}, \quad
	\tilde{r}^{-}_{j}=(\epsilon^{-}_{j}-\lambda) \tilde{x}^{-}_{j},
	\label{}
\end{equation}
implying that the error in the residual $\tilde{r}_i^+$ (resp. $\tilde{r}^{-}_{j}$)
is amplified by $\epsilon^{+}_{i}-\lambda$ (resp. $\epsilon^{-}_{j}-\lambda$) from
$\tilde{x}^{+}_{i}$ (resp. $\tilde{x}^{-}_{j}$).
In order to quantify the relative amplification of the error projected
to the negative energy band in the residual $r$ compared to the
amplification of the error projected to the positve energy band,
we define the
\textit{amplification factor} as follows
\begin{equation}
	\gamma_{\mathrm{LOBPCG}} = \max_{\lambda}
	\frac{\max_{j}\abs{\epsilon^{-}_{j}-\lambda}}{\max_{i}\abs{\epsilon^{+}_{i}-\lambda}}.
	\label{}
\end{equation}
Since $\lambda$ is a desired eigenvalue contained in the positive energy band,
the assumption \eqref{eqn:assumptionDW} implies further that
\begin{equation}
	\abs{\epsilon^{-}_{j}-\lambda}\sim \Delta,\quad
	\abs{\epsilon^{+}_{i}-\lambda}\sim w.
	\label{}
\end{equation}
In consequence, we have
\begin{equation}
	\gamma_{\mathrm{LOBPCG}}\sim \frac{\Delta}{w},
	\label{}
\end{equation}
and the relative error projected to the negative energy band in
$r$ is amplified by a factor ${\Delta}/{w}$ compared to that in $x$.
The error remains in the preconditioned residual $y$
and propagates into the subspace used in the next LOBPCG iteration (line 11 in
Alg.~\ref{alg:LOBPCGF}).
In case of $\frac{\Delta}{w} = 300$,
even if the initial error is small, say $\max_{j}|\tilde{x}^{-}_{j}|\sim 10^{-15}$,
only after $6$ iterations, the error of the eigenfunctions
projected to the negative energy band will accumulate to
$\mathcal{O}(1)$ in the worst case scenario.

In order to compensate for the amplified error projected to the negative energy
band, the LOBPCG-F method applies a filtering step to the
preconditioned residual $Y$ before the error propagates to the next
LOBPCG iteration (line 10 in Alg.~\ref{alg:LOBPCGF}).  The filtering
function takes the form
\begin{equation}
	f(E) = \frac{1}{\Omega^2}(E-C)^2,
	\label{eqn:filter}
\end{equation}
where the shift constant $C$ is chosen to be in the
negative energy continuum (see Fig.~\ref{fig:schemeLOBPCG}),
and $\Omega$ is a normalization constant so that
$f(E) = 1$ when $E$ is the largest eigenvalue in the positive energy
band.  We remark that although filtering functions of other forms can
also be employed, the quadartic filtering function~\eqref{eqn:filter}
requires only $2$ extra matrix-vector multiplications, and achieves
nearly the minimum cost.  Under the assumption~\eqref{eqn:assumptionDW},
we have $\Omega\sim \Delta$. In the following we show how the filter
\eqref{eqn:filter} works.

Again using the eigen-decomposition for $y$ after the filtering step
(line 10 in Alg.~\ref{alg:LOBPCGF}) and assuming the identity preconditioner $T$
lead to
\begin{equation}
	y = \sum_{i} \psi^{+}_{i}\tilde{y}^{+}_{i}
	   +\sum_{j} \psi^{-}_{j}\tilde{y}^{-}_{j},
	\label{}
\end{equation}
where
\begin{equation}
	\tilde{y}^{+}_{i}=f(\epsilon_{i}^{+})(\epsilon^{+}_{i}-\lambda) \tilde{x}^{+}_{i}, \quad
	\tilde{y}^{-}_{j}=f(\epsilon_{j}^{-})(\epsilon^{-}_{j}-\lambda) \tilde{x}^{-}_{j}.
	\label{}
\end{equation}
The amplification factor of the LOBPCG-F method becomes
\begin{equation}
	\gamma_{\mathrm{LOBPCG-F}} = \max_{\lambda}
	\frac{\max_{j}\abs{f(\epsilon^{-}_{j})(\epsilon^{-}_{j}-\lambda)}}{
	\max_{i}\abs{f(\epsilon_{i}^{+})(\epsilon^{+}_{i}-\lambda)}}.
	\label{}
\end{equation}
Again under the assumption~\eqref{eqn:assumptionDW}, we have
\begin{equation}
	\abs{f(\epsilon^{-}_{j})(\epsilon^{-}_{j}-\lambda)}\sim
	\frac{w^2\Delta}{\Omega^{2}},\quad
	\abs{f(\epsilon^{+}_{i})(\epsilon^{+}_{i}-\lambda)}\sim
	\frac{\Delta^2 w}{\Omega^{2}},
	\label{}
\end{equation}
and then
\begin{equation}
	\gamma_{\mathrm{LOBPCG-F}}\sim \frac{w}{\Delta}.
	\label{}
\end{equation}
Therefore after each filtering step,
the relative error projected to the negative energy band in $y$ is reduced by
a factor $w/\Delta$ compared to the relative error in $x$.  As mentioned
in Section~\ref{sec:intro}, the spectral radius of the matrix originated
from the spectrum folding type methods~\cite{WangZunger1994}
deteriorates as $\Delta^2$  when $\Delta$ increases to infinity, and
an efficient iterative method is difficult to be designed for the resulting
matrix eigenvalue problem. On the contrary,
the efficiency of the LOBPCG-F method improves as $\Delta$ increases,
characterized by the factor ${w}/{\Delta}$.

\begin{figure}[htpb]
  \begin{center}
    \subfloat[]{\includegraphics[width=0.35\textwidth]{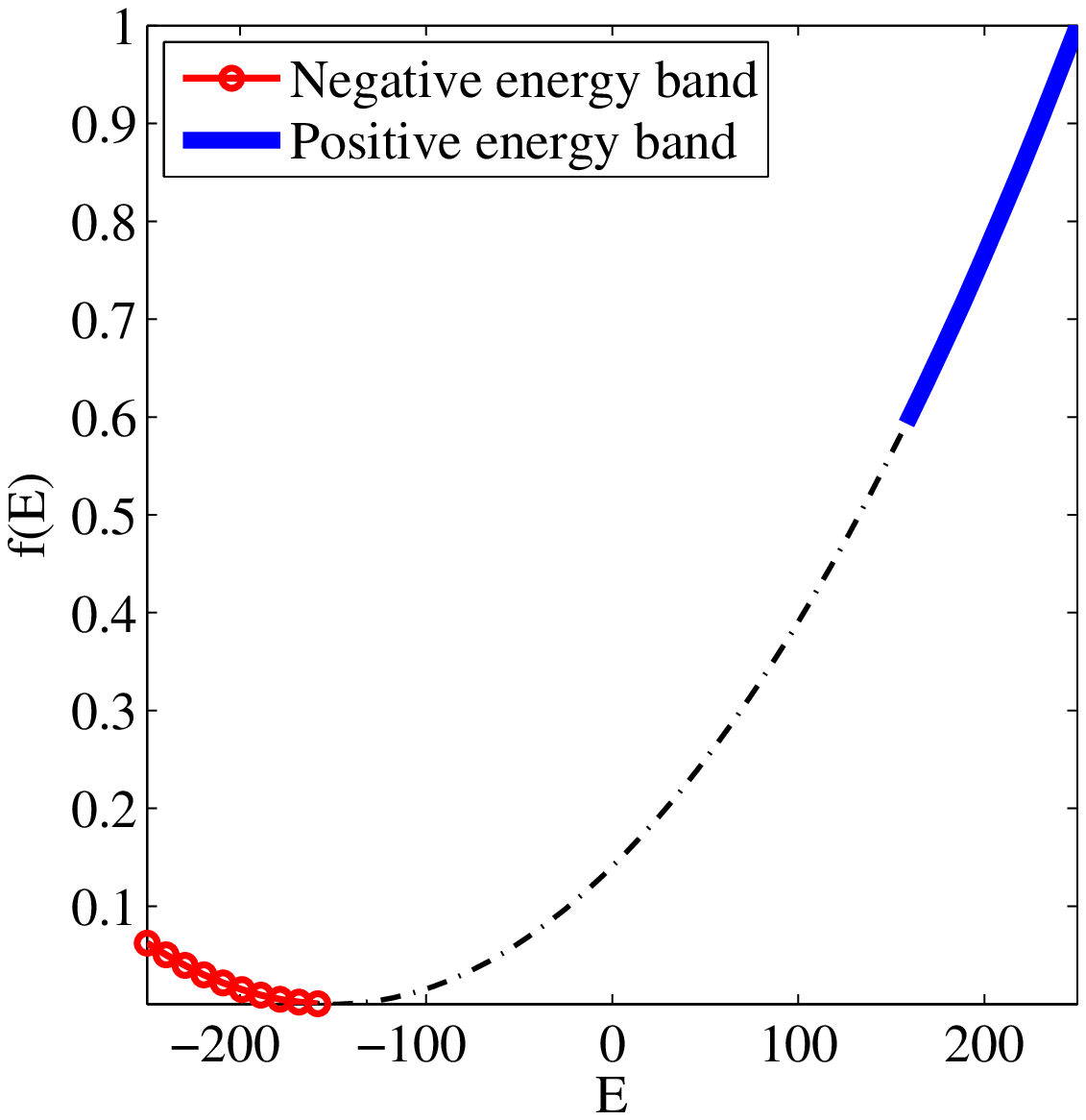}}
    \subfloat[]{\includegraphics[width=0.35\textwidth]{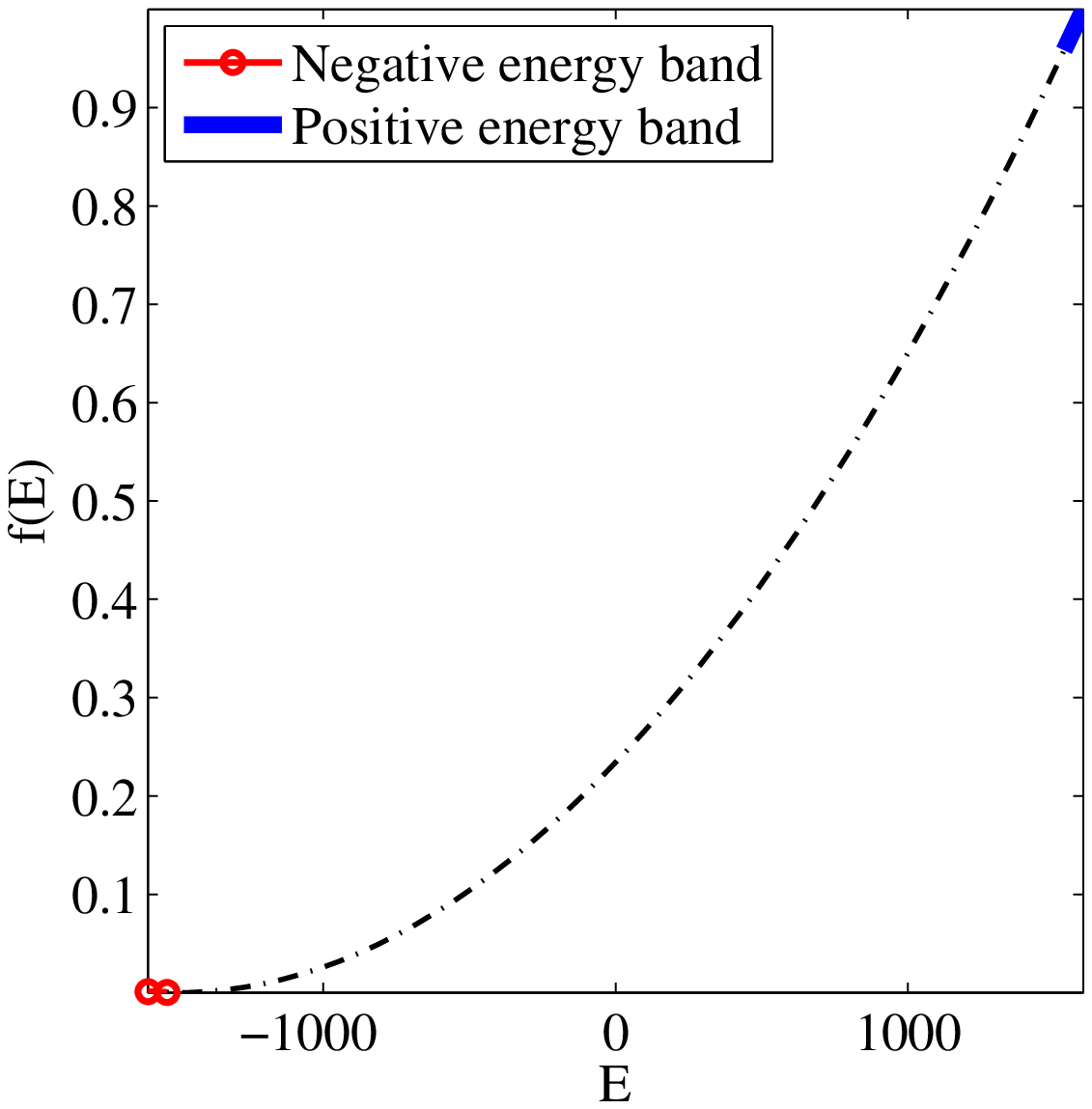}}
  \end{center}
	\caption{The filtering function $f(E)$ for $w=100$ au with a small
	gap $\Delta=300$ au (a) and with a large gap $\Delta=3000$ au (b).  The function values
	correspond to the positive energy band (blue solid line) and the
	negative energy band (red circles) are separated
	by the function values in the spectral gap (black dotted line).  }
  \label{fig:schemefandfE}
\end{figure}

We plot the shape of the filtering function \eqref{eqn:filter} for the case $w=100$ au with
a very small gap $\Delta=300$ au (correspondingly $c\approx 12$ au) in Fig.~\ref{fig:schemefandfE} (a)
and for $w=100$ au with a larger gap $\Delta=3000$ au (correspondingly
$c\approx 39$ au) in Fig.~\ref{fig:schemefandfE} (b).
We observe that even for such relatively small gaps, the filtering
function already quickly reduces the error in the negative energy band.
The filtering function \eqref{eqn:filter} is increasingly more effective
as the gap $\Delta$ increases.  In the practical calculation, the gap
$\Delta\approx 37000$ au ($c\approx 137$ au), and thus the filtering
function can effectively control the error of the eigenfunctions
projected to the negative energy band in each LOBPCG-F iteration.

Finally we need to address the initial condition such that
$\max_{j}|\tilde{x}^{-}_{j}|$ is small.   This can be achieved by the
lines $1$--$3$ in Alg.~\ref{alg:LOBPCGF} using $N_{\mathrm{init}}$ steps
of the same filtering function.  A high quality input of the initial vectors
$X_{0}$ can help reduce the number $N_{\mathrm{init}}$.
In the DKS
calculation, a good set of initial vectors can be obtained by using the
wavefunctions obtained from the KS or the KSSO calculations.
The converged KS or KSSO wavefunctions can be used as
the initial guess for the large component of the DKS spinors,
and the initial guess for the small component of the DKS spinors
can be obtained by applying the kinetic balance prescription \eqref{RKB} to them.
The good
input of the initial vectors is also readily available in consecutive
SCF steps, in which $X_{0}$ simply takes the output $X$ in the previous
SCF step, and the error projected to the negative energy band in $X$ is
negligible as controlled by the LOBPCG-F method.  Using such strategy,
we find that the LOBPCG-F algorithm is robust by setting
$N_{\mathrm{init}}=3$ for the first SCF iteration (for safety), and then
setting $N_{\mathrm{init}}=1$ in the following SCF steps.

\subsection{Implementation}

Compared to the standard LOBPCG algorithm which applies the Hamiltonian
matrix once per iteration step, the LOBPCG-F algorithm applies the
Hamiltonian matrix for $2$ extra times per iteration due to the
filtering step \eqref{eqn:filter}.  The computational cost of the numerical linear algebra
operations, such as for the orthogonalization step and for the
projected Rayleigh-Ritz eigenvalue problem, remains to be the same as that in the
standard LOBPCG algorithm. Furthermore, the implementation of LOBPCG-F
algorithm is very simple. Since the implementation of the LOBPCG
algorithm always provides an interface for the preconditioner, one only
needs to apply the filtering function after the standard preconditioner to
eliminate the negative energy states components as shown in Alg.~\ref{alg:LOBPCGF},
and the overall
framework for LOBPCG, such as that provided by
BLOPEX~\cite{KnyazevArgentatiLashukEtAl2007} does not need to be
changed.

\section{Numerical results}
\label{sec:numericalresults}

We implement the LOBPCG-F method for solving the DKS system \eqref{vDKS} using the HGH
pseudopotential~\cite{HartwigsenGoedeckerHutter1998}, with the local and
nonlocal pseudopotential implemented fully in the real
space~\cite{PaskSterne2005}.  The exchange-correlation functional
under local density approximation
(LDA) is used.  Together with the
module DKS for solving the four-component DKS system, we also implement
the standard LOBPCG method for solving the KS density functional theory
(module KS), and for solving the
two-component KS density functional theory with SOC in the
pseudopotential framework (module KSSO).
Although we implement the modules from scratch, the LOBPCG-F method can
also be implemented without too much efforts by developers of other packages
such as ABINIT~\cite{Abinit1}, Quantum
ESPRESSO~\cite{GiannozziBaroniBoniniEtAl2009} \etc.
The major difference between
the module KS and KSSO is that KSSO includes the spin-orbit coupling
term between the core electrons and the valence electrons in the form of  $\sum_{l}
V_{l}^{\text{SO}}(\vr,\vr') \vL'\cdot\vS$ as in Eq.~\eqref{eqn:HGH}.
The real and complex arithmetics are equally handled in our
implementation.  The description of the modules KS, KSSO and DKS is
summarized in Table~\ref{tab:code}.
Since our main focus is the
relativistic effect such as SOC, the module KS directly
uses a two-component spinor rather than a one-component spinor.
However, this does not lead to changes in the computed physical
quantities such as the total energy.  The availability of modules KS and
KSSO allows us to directly benchmark our implementation with existing
software such as ABINIT for electronic structure
calculation. To facilitate the presentation of the numerical results,
we adopt the standard convention that all energies are shifted by
$-mc^2$, so that the positive energy continuum starts from $0$ rather
than $+mc^2$.

\begin{table}[htpb]
	\centering
	\caption{The number of components for describing a spinor
	($N_{\text{spinor}}$), the arithmetic, and the
	diagonalization solver for modules KS, KSSO and DKS in our
	implementation.}
	\label{tab:code}
	\begin{tabular}{l|l|l|l|l}
    \toprule
		Module & $N_{\text{spinor}}$ & Arithmetic & Solver& Description\\
		\midrule
		KS     & 2  & Real & LOBPCG &Kohn-Sham\\
		KSSO   & 2  & Complex & LOBPCG& Kohn-Sham with \\
		       &    &         &       & spin-orbit coupling \\
		DKS    & 4  & Complex & LOBPCG-F& Dirac-Kohn-Sham \\
		\bottomrule
	\end{tabular}
\end{table}

More computational details of our implementation and the numerical
results are as follows.  The preconditioner proposed by Teter
\etal~\cite{PayneTeterAllenEtAl1992,TeterPayneAllan1989} is employed by
both LOBPCG and LOBPCG-F methods.  Anderson mixing~\cite{Anderson1965}
with Kerker preconditioner~\cite{Kerker1981} is used for the SCF
iteration.  Gamma point Brillouin sampling is used for simplicity for
all calculations, even for crystal systems such as Bi$_{2}$Se$_{3}$.
This is because the purpose of this manuscript is to demonstrate the
capability of the LOBPCG-F method for solving DKS systems.  The support
of $k$-point sampling will be added in the future work.  The density and
the wavefunction are resolved using the same grid in both real space and
Fourier space, and the grid size is measured by the corresponding
kinetic energy cutoff in the Fourier space in atomic unit, denoted by
$E_{\text{cut}}$. All computational experiments are performed on the
Hopper system at the National Energy Research Scientific Computing
(NERSC) center.  Each Hopper node consists of two twelve-core AMD
``MagnyCours'' 2.1-GHz processors and has 32 gigabytes (GB) DDR3
1333-MHz memory. Each core processor has 64 kilobytes (KB) L1 cache and
512KB L2 cache.  All modules KS, KSSO and DKS are implemented
sequentially, and one core processor is used in each computation.

\subsection{Setup}

\begin{figure}[htpb]
  \begin{center}
    \includegraphics[width=0.6\textwidth]{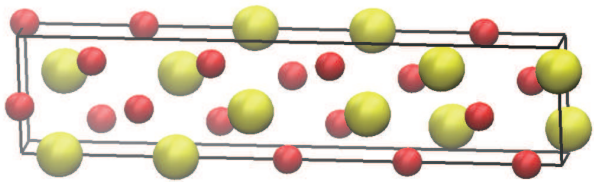}
  \end{center}
  \caption{The converted orthorhombic supercell for describing
  a topological insulating system Bi$_{2}$Se$_{3}$ with $12$ Bi atoms
  (large yellow balls) and $18$ Se atoms (small red balls).}
  \label{fig:Bi2Se3cell}
\end{figure}

We present numerical results for computing the total free energy of
three dimers systems Pt$_{2}$, Au$_{2}$, and TlF, as well as a condensed
matter system Bi$_{2}$Se$_{3}$. For simplicity of demonstration, the
electronic structure calculation of the dimers are not computed with
their optimal bond lengths.  Instead, all the dimers are placed in a
supercell of dimension $15.000$ au, $10.000$ au, $10.000$ au along the
$x, y, z$ directions respectively. The positions of the dimers are placed
at $(5.000, 0.000, 0.000)$ au and $(10.000, 0.000, 0.000)$ au,
respectively.  In the HGH pseudopotential, the elements Pt, Au and Tl
have the option of including semicore
electrons~\cite{HartwigsenGoedeckerHutter1998} or not, and our implementation
can handle both cases.  Here we treat Pt atoms
without semicore electrons, and treat Au atoms and Tl atoms with
semicore electrons.  Our implementation assumes an
orthorhombic supercell.  Non-orthorhombic cells can be converted into
orthorhombic cells for the total energy computation. For example, the
topological insulator phase of Bi$_{2}$Se$_{3}$ has rhombohedral crystal
structure~\cite{ZhangLiuQiEtAl2009}. The rhombohedral structure can be converted
into an orthorhombic cell as shown in Fig.~\ref{fig:Bi2Se3cell} with
$12$ Bi atoms and $18$ Se atoms in the supercell. The lengths of the
converted orthorhombic supercell is $7.820$ au, $13.544$ au, and $54.123$ au
along the $x,y,z$ directions, respectively.  The kinetic energy cutoff
$E_{\text{cut}}$ is set to be the same in ABINIT and in our
implementation, and $E_{\text{cut}}$ is high enough so that the
difference in the total free energy per atom is less than $1$ meV.
Neither Bi nor Se allows
semicore treatment in HGH pseudopotential.  More details of the setup of
the computational systems are illustrated in Table~\ref{tab:setup}.
Note that TlF requires particularly large kinetic energy cutoff. This
is mainly due to localized pseudopotential of the F atom, which
requires a very fine numerical grid to resolve.

\begin{table}
	\centering
\caption{The kinetic energy cutoff $E_{\text{cut}}$, the number of
	valance electrons and the treatment of semicore electrons for the systems
	under consideration.  The same kinetic energy cutoff is used in both
	ABINIT and our implementation.}
	\label{tab:setup}
	\begin{tabular}{c|c|c|c}
		\toprule
		System & $E_{\text{cut}}$ & \# Electron & Semicore \\
		\midrule
		Pt$_{2}$ & 50 &  20 & No  \\
		Au$_{2}$ & 50 &  22 & Yes  \\
		TlF      & 400 &  20 & Yes  \\
		Bi$_{2}$Se$_{3}$ & 45 & 168 & No\\
    \bottomrule
	\end{tabular}	
\end{table}

\subsection{Calibration with ABINIT}

We compare our result with ABINIT~\cite{Abinit1} which also supports the
usage of the HGH pseudopotential and the LDA approximation. ABINIT has the
capability of performing KS calculation (by setting
\textsf{nspinor=1,nspden=1}) and KSSO calculation with vector
magnetization of the spin density (by setting \textsf{nspinor=2,nspden=4}).
At present, ABINIT does not provide the DKS module.
As shown by other researchers, the SCF iteration
can become oscillatory for spin-unrestricted calculations, which is
known as the ``spin-sloshing'' and exists even for a simple
spin-unrestricted O$_2$ dimer system~\cite{MarksLuke2008}.
In order to accelerate the SCF iteration,
the temperature is set to be $5000$K in all simulations reported below.
Although the increase of the temperature in the outer SCF loop directly
affects the convergence of density, magnetization and potential,
it does not directly affect the
convergence behavior of individual eigenfunctions, which is controlled
by eigensolvers in the inner loop such as LOBPCG or LOBPCG-F.
Finite temperature formulation of the KS
density functional theory~\cite{Mermin1965} is used, and the total free
energy
is the quantity to be measured~\cite{AlaviKohanoffParrinelloEtAl1994}.
In particular, the total free energy computed from modules KS, KSSO and
DKS in our implementation are denoted by $E_{\text{KS}},
E_{\text{KSSO}}, E_{\text{DKS}}$, respectively,
while those computed from the corresponding modules KS and KSSO in ABINIT
are denoted by $E^{\text{ABINIT}}_{\text{KS}}, E^{\text{ABINIT}}_{\text{KSSO}}$,
respectively. To facilitate the illustration, all the energies will be
measured in the unit of au, and all the differences of energies
will be measured in the unit of meV ($1 \text{au} \approx 27211
\text{meV}$ ).

\begin{table}[htbp]
  \centering
  \caption{The total free energies $E^{\text{ABINIT}}_{\text{KS}}$
	calculated from ABINIT, and $E_{\text{KS}}$ from module KS in our
	implementation for systems under consideration.}
	\label{tab:accuracyKS}%
    \begin{tabular}{c|c|c|c}
    \toprule
		System & $E^{\text{ABINIT}}_{\text{KS}}$ & $E_{\text{KS}}$ &
		$(E_{\text{KS}}-E^{\text{ABINIT}}_{\text{KS}})/N_{\text{atom}}$\\
           & au    & au    & meV \\
    \midrule
    Pt$_2$   & -52.364236 & -52.364274 & -0.51 \\
    Au$_2$   & -66.438610 & -66.438601 & 0.12 \\
    TlF      & -74.156822 & -74.156828 & -0.08 \\
    Bi$_{2}$Se$_{3}$ & -237.357221 & -237.357063 & 0.14 \\
    \bottomrule
    \end{tabular}%
\end{table}%

In Table~\ref{tab:accuracyKS},
we compare the total free energies for the KS calculation.
The differences of energies per atom are less than $1$
meV in all cases.
In Table~\ref{tab:accuracyKSSO}, we compare the total free energies
for the KSSO calculation.
Similarly, the differences of energies per atom are also less than $1$ meV in all
cases.  Tables~\ref{tab:accuracyKS} and \ref{tab:accuracyKSSO} indicate
that the KS and KSSO modules in our implementation are accurate.
Comparing the data shown in Tables~\ref{tab:accuracyKS} and \ref{tab:accuracyKSSO},
we find
that the total free energies are consistently lower when SOC effect is
considered.

\begin{table}[htbp]
  \centering
  \caption{The total free energies $E^{\text{ABINIT}}_{\text{KSSO}}$
	calculated from ABINIT, and $E_{\text{KSSO}}$ from module KSSO in our
	implementation for systems under consideration.}
  \label{tab:accuracyKSSO}%
    \begin{tabular}{c|c|c|c}
    \toprule
		System & $E^{\text{ABINIT}}_{\text{KSSO}}$ & $E_{\text{KSSO}}$ &
		$(E_{\text{KSSO}}-E^{\text{ABINIT}}_{\text{KSSO}})/N_{\text{atom}}$\\
           & au    & au    & mev \\
    \midrule
    Pt$_2$            & -52.411523 & -52.411564 & -0.56 \\
    Au$_2$            & -66.473300 & -66.473302 & -0.03 \\
    TlF               & -74.178538 & -74.178602 & -0.87 \\
    Bi$_{2}$Se$_{3}$  & -237.728372     & -237.728987 & -0.56 \\
    \bottomrule
    \end{tabular}%
\end{table}%

\subsection{DKS results}

Table~\ref{tab:resultDKS} demonstrates the total free energies computed
using the module DKS in our implementation, for which ABINIT does not
provide the same functionality,
and shows the differences of the energies per atom between calculations using modules
KS, KSSO and DKS.  The energies obtained from DKS is systematically
lower than those in KS and KSSO.  The full relativistic correction
described by DKS is large (in the order of hundreds of meV per atom)
compared to the result obtained by KS which only takes into account the
scalar relativistic effect in the level of pseudopotential.  Most of the
correction originates from SOC effect.
The remaining correction due to the difference between the Dirac
description (DKS) and the Pauli description (KSSO, with the relativistic
effect taken account only in the HGH pseudopotential) is generally two
orders of magnitude smaller than the SOC effect.
Furthermore, since the relativistic effect of the core electrons is more
significant than that of the valence electrons, the difference between
DKS and KSSO is particularly small when the semicore treatment is not
present, such as in the case of Pt$_2$ and Bi$_{2}$Se$_{3}$.
With the same HGH pseudopotential~\eqref{eqn:HGH}, the correction from
the Dirac description compared to the Pauli description is found to be
smaller than the transferability error of the pseudopotentials.
Our study also agrees
with the recent study using the planewave implementation of the ZORA
equations -- one kind of approximate two-component relativistic theory
by NWChem~\cite{Nichols2009}.
We also note that in TlF, the difference
between DKS and KSSO is around $10\%$ of the effect due to SOC.
To the extent of our knowledge, our result for the first time reveals
the quantitative difference between the Pauli description and the full
Dirac description of relativistic effects for the valence electrons in
the pseudopotential framework for the systems under study.

\begin{table}[htbp]
  \centering
  \caption{The total free energies  $E_{\text{KSSO}}$ from module DKS in our
	implementation for systems under consideration.}
  \label{tab:resultDKS}%
    \begin{tabular}{c|c|c|c}
    \toprule
    System & $E_{\text{DKS}}$ &
		$(E_{\text{DKS}}-E_{\text{KS}})/N_{\text{atom}}$ &
		$(E_{\text{DKS}}-E_{\text{KSSO}})/N_{\text{atom}}$ \\
		       & au & meV & meV \\
    \midrule
    Pt$_2$            & -52.411595 & -643.82 & -0.41 \\
    Au$_2$            & -66.473702 & -477.56 & -5.44 \\
    TlF               & -74.181219 & -331.85 & -35.60 \\
    Bi$_{2}$Se$_{3}$  & -237.730530 & -338.75 & -1.40 \\
    \bottomrule
    \end{tabular}%
\end{table}%

\subsection{Computational efficiency}

We demonstrate the computational efficiency of the LOBPCG-F method in
terms of the wall clock time per SCF iteration for Pt$_{2}$, Au$_{2}$
and TlF in Table~\ref{tab:efficiency}. Each SCF iteration consists of
$3$ LOBPCG iterations for the KS and the KSSO calculation, and $3$
LOBPCG-F iterations for the DKS calculation.  The computation of
Bi$_{2}$Se$_{3}$ uses significant amount of virtual memory due to the
large number of valance electrons in the system, and the corresponding
computational time is therefore not meaningful and not reported here.  This
will not be a problem when our parallel implementation of our method is
introduced in the future.  We also remark that compared to the standard
LOBPCG method, the LOBPCG-F method does not introduce any additional
memory cost.

From KS to KSSO the computational time increased by a factor of $5\sim
6$.  As discussed before, the module KS uses a two-component
formulation, and the increase of the computational time is mostly due to
the change from real to complex arithmetic. For computing the same
quantity, the complex arithmetic is $4$ times more expensive than the
real arithmetic.  The remaining difference comes from that KSSO uses
complex to complex (c2c) Fourier transform, and KS uses real to complex
(r2c) and complex to real (c2r) Fourier transform.

From KSSO to DKS the computational time increases by a factor around
$4$.  In this process, the change of the number of components from $2$
to $4$ inevitably increases the computational time by a factor of $2$.
The remaining factor of $2$ in the increased computational time
originates from the difference between the LOBPCG and the LOBPCG-F
method.  Each LOBPCG iteration applies the Hamiltonian to all spinors
once, and each LOBPCG-F iteration applies the Hamiltonian operator to
all spinors for $3$ times to filter the components in the negative
energy band.  However, since the linear algebra operations (such as the
orthogonalization procedure of the spinors) are the same in the LOBPCG
and the LOBPCG-F method, the usage of LOBPCG-F only increases the
computational time by a factor around $2$ in practice.

\begin{table}[htbp]
  \centering
  \caption{The wall clock time (in the unit of sec) for performing one
	step of SCF iteration for systems under consideration.}
  \label{tab:efficiency}%
    \begin{tabular}{c|c|c|c}
    \toprule
    System & KS  & KSSO  & DKS \\
    \midrule
    Pt$_2$ & 4     & 26    & 113 \\
    Au$_2$ & 6     & 37    & 155 \\
    TlF    & 148   & 709   & 2787 \\
    \bottomrule
    \end{tabular}%
\end{table}%

\begin{figure}[htpb]
	\begin{center}
		\includegraphics[width=0.5\textwidth]{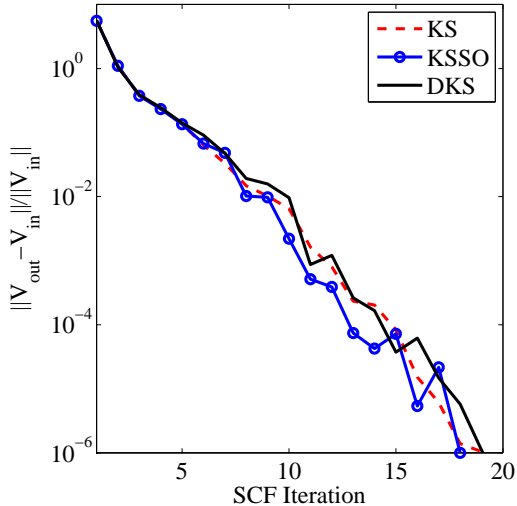}
	\end{center}
	\caption{Convergence of the SCF iteration for KS (red dashed line),
	KSSO (blue line with circles) and DKS calculations (black solid line)
	for Au$_{2}$.}
	\label{fig:scfconv}
\end{figure}

Fig.~\ref{fig:scfconv} compares the convergence of the SCF
iteration for KS, KSSO and DKS using Au$_{2}$ as an example. The convergence of
the SCF iteration is measured by the quantity
$\norm{V_{\text{out}}-V_{\text{in}}}/\norm{V_{\text{in}}}$, where
$V_{\text{in}}$ is the local part of the effective potential before each
SCF iteration, and $V_{\text{out}}$ is the effective potential after
each SCF iteration.  Fig.~\ref{fig:scfconv} indicates that the usage of
the LOBPCG-F method does not deteriorate the convergence rate of the SCF
iteration.  Combining Table~\ref{tab:efficiency} and
Fig.~\ref{fig:scfconv}, we conclude that the LOBPCG-F is a robust
and efficient method for solving the DKS system in practice.

\section{Conclusion and outlook}
\label{sec:conclusion}

In this manuscript we develop for the first time a simple and efficient
iterative algorithm, named the LOBPCG-F method, for directly solving the
Dirac-Kohn-Sham (DKS) equations in the relativistic density functional
theory.  By adding an additional filtering step in the preconditioning
stage of the LOBPCG method, the LOBPCG-F method is able to compute the
desired eigenvalues and eigenvectors in the positive energy band without
computing any state in the negative energy band.  The LOBPCG-F method
requires only $2$ extra computational costs and little extra coding
effort, and thus remarkably facilitates the transition from
nonrelativistic Kohn-Sham (KS) calculations using LOBPCG methods to
relativistic DKS calculations in studying the relativistic effect such
as spin-orbit coupling, without the need of approximating the DKS
equation by two-component relativistic theory.  We also remark that even
though the LOBPCG-F method is efficient for solving the DKS systems, it
is still significantly more expensive to solve the more complicated
four-component DKS systems compared to the KS systems or the KSSO
systems. In practice one should decide whether two-component theory or
the four-component theory should be pursued by balancing the desired
efficiency and accuracy.

The efficiency of the filtering step is determined by the factor $w/\Delta$,
where $\Delta$ is the spectral gap between the positive
and the negative energy band, and $w$ is the maximum of the widths of
the positive and the negative energy band (see Fig.~\ref{fig:schemeLOBPCG}).
The smaller the factor is, the more efficiently the filter performs.
It is worth highlighting that the proposed filtering technique is more general and can be
embedded into other preconditioned iterative solvers,
while the LOBPCG method is employed in this work for its high efficiency
in the electronic structure calculation.
We demonstrate the
applicability of the LOBPCG-F method in the pseudopotential framework in
the planewave basis set, in which the condition $\Delta\gg w$ is
satisfied.  The planewave basis set automatically satisfies the kinetic
balance prescription and is free from the variational collapse.  Our
results compared with ABINIT indicate that our implementation
is accurate, and the LOBPCG-F method does not lead to deterioration in the
convergence rate of the SCF iteration.  We directly observe that the
difference between the total energies between the two-component KS density functional
theory with SOC description (module KSSO) and
the DKS description (module DKS) is no more than a few tens of meV per atom for
systems under study, and thus it is not cost-effective to solve the
four-component DKS problem in the pseudopotential framework for most systems
in practice.  The LOBPCG-F method will be more effective for
studying the relativistic effects in solving the all-electron DKS
system directly when the condition $\Delta\gg w$ is satisfied, such as
using the linearized augmented plane-wave basis set (LAPW) as in the
WIEN2k
software, or the
localized basis set given that the basis set remains well conditioned.
This will be our future work.

\vspace{1em}
\noindent{\bf Acknowledgment:}

This work is partially supported by the Laboratory Directed Research and
Development Program of Lawrence Berkeley National Laboratory under the
U.S. Department of Energy contract number DE-AC02-05CH11231 (L.~L.), and
the National Natural Science Foundation of China under the grant number
11101011 and the Specialized Research Fund for the Doctoral Program of
Higher Education under the grant number 20110001120112 (S.~S.).  S.~S.
acknowledges the support from the Program in Applied and Computational
Mathematics at Princeton University and from the Lawrence Berkeley
National Laboratory for his sabbatical visit in the first half of 2012,
during which the work on this paper is initiated.  The authors are
grateful to the useful discussions with Roberto Car, Wibe De Jong,
Jianfeng Lu, Emil Prodan, Chao Yang, and Yong Zhang.

\end{document}